\documentclass{JHEP}

\usepackage{amssymb}

\usepackage{amsfonts}

\usepackage{amsbsy}

\usepackage{epsfig}

\newcommand{\im}{\mathop{\rm Im}\nolimits}

\renewcommand{\cal}{\mathcal}

\def\C{{\Bbb C}}

\def\P{{\Bbb P}}

\def\Z{{\Bbb Z}}

\def\s*{\boldsymbol{*}}

\title{D-Branes and Vanishing Cycles in Higher Dimensions.}

\author{Mark Raugas\\ Department of Mathematics, Columbia University\\
New York, NY 10027, USA\\ \email{raugasm@math.columbia.edu}}

\abstract{We investigate the quantum volume of D-branes wrapped around
cycles of various dimension in Calabi-Yau fourfolds and fivefolds.  Examining the cases of the sextic and heptic hypersurface Calabi-Yau varieties, as well as one example in weighted projective space, we find expressions for periods which vanish at the singular point analogous to the conifold point.  As in the known three-dimensional cases, it is the top dimensional cycle which attains zero quantum volume, even though lower dimensional cycles remain non-degenerate, indicating this phenomena to be a general feature of quantum geometry.}

\begin{document}

\section{Introduction}

Quantum geometry often exhibits marked differences from the expectations
one derives from classical intuitions.  These novel features have in recent years helped aid the understanding of many phenomena in string theory \cite{small_distances1, quantum_volumes}.  For example, worldsheet instanton effects lead one to determine that it is the collection of all rational curves over a given Calabi-Yau manifold which can be thought of as the quantum version of the variety \cite{morrison_quintic}.  In addition, by using nonperturbative probes such as D-Branes, one discovers that one must also consider supersymmetric cycles of all dimensions as relevant to the quantum geometry of a variety \cite{D_geometry}.  The idea of quantum volume has been studied for a large class of Calabi-Yau threefolds; determining the states which vanish along the discriminant locus and exhibiting the emergence of a massless degree of freedom quite generically \cite{S, GL1}.  More recently, for the Quintic, much work has been done to study the BPS spectrum of states in great detail, making predictions based on both boundary conformal field theory methods and $\cal{N}=2$ supergravity approaches \cite{boundary_states, Ishibashi, Douglas_quintic, Moore_arithmetics, denef, spectrum}.  

In this paper, we concentrate our efforts to extending the above to include the class of examples which are higher dimensional Calabi-Yau manifolds.  The technology of mirror symmetry was first extended to higher dimensions in the paper \cite{higher}.  We use those results, combined with the advances made in the works listed above, to determine if analagous statements about vanishing cycles can be made in the case of Calabi-Yau manifolds of higher dimension.  

First we review the ideas of quantum volume and the technology of Meijer periods developed in \cite{GL1}, and generalize them to the case of hypersurface Calabi-Yau manifolds in higher dimensions.  We then analyze the cases of the sextic hypersurface in $\P^5$ and the heptic hypersurface in $\P^6$.  We then consider an example in a  weighted projective space.  For each of the three examples, we obtain simple expressions for collapsing cycles which vanish along the discriminant locus.  We end with some conclusions and remarks, as well as some directions for further research.

%%%%%%%%%%%%%%%%%%%%%%%%%%%%%%%%%%%%%%%%%%%%%%%%%%%
\section{Higher Dimensional Calabi-Yau Manifolds}

\subsection{Quantum Volumes}

As is usual, we may define the \emph{quantum volume} of a holomorphic even dimensional
cycle in a $d$-dimensional algebraic variety with trivial anticanonical bundle to
be equal to the quantum corrected mass of the (IIA) BPS saturated
D-brane state wrapping it; this is equal to the \emph{classical}
mass of its mirror $d$-cycle.  

This agrees with our naive notion of volume in the classical regime, but as we move in moduli space, 
corrections arise which  alter these volumes as functions of the moduli.  In this manner, we obtain an 
exact stringy expression for the volume of a given even dimensional holomorphic cycle $\gamma$ in a 
Calabi-Yau $d$-fold $X$ with in terms of the normalized period of its mirror $d$-cycle $\Gamma$:

 \begin{equation}
  \label{BPSmassesinIIB} M(\gamma)= |Z(\Gamma)| =
  \frac{|\int_{\Gamma}{\Omega(z)}|}{(\int_{Y}{i \, \Omega(z)\wedge
  {\overline \Omega(z)}} )^{1/2}}~~=
  \frac{|q^i\int_{\Gamma_i}{\Omega}|}{(\int_{Y}{i \, \Omega(z)\wedge
  {\overline \Omega(z)}})^{1/2}} ~~.
 \end{equation}

In the above, $\{\Gamma_i \}_i$, $i=1,\ldots,h^{d}(Y)$, is
an integral basis of $H_{d}(Y)$, with $h^{d}(Y)$ being the sum of the ranks of the $d$-dimensional Dolbeault cohomology.  $\Omega (z)$ is the holomorphic
$d$-form, written with explicit dependence on the moduli $z$.  The 
$q^i$ are the integral charges of the cycle with respect to the
$\Gamma_i$, and $\int_{\Gamma_i}{\Omega(z)}$ are the periods of
the holomorphic $d$-form.

In this paper, we consider varieties $X$ with $h^{1,1}(X)=1$.  As such, we may label  the point $z=0$ as the large complex structure limit, i.e. the complex structure for the mirror  variety $Y$ which is mirror to the large volume limit of $X$.  In this class of examples, holomorphic cycles of real dimension $2j$ on $X$ are mirror to $d$-cycles on $Y$.

\subsection{Meijer Periods}

The technology of Meijer periods \cite{Meijer_refs, Norlund,GL1}
allows us to write down a basis of solutions to the Picard-Fuchs
equation associated to a given variety $Y$, which is indexed by
the leading logarithmic behavior of each solution.  

This is useful, as D-Branes who wrap $2j$ dimensional cycles on $X$ will be mirror to $d$-dimensional branes on $Y$ which have periods with leading ${\rm log}^j z$ behavior near $z = 0$.  Thus, finding a complete set of periods of $\Omega(z)$ and classifying their leading logarithmic behavior gives us a means of identifying
the dimension of their even-cycle counterpart on $X$.  As usual in this context, when we speak about a cycle on $X$ of real dimension $2j$, we refer to a cycle on $X$ with $2j$ being the maximal dimensional component, but with the identity of any lower dimensional ``dissolved'' cycles left unspecified.  

In particular, in the case of higher dimensional Calabi-Yau manifolds, we are still able to 
find a basis of solutions, each representing a single BPS $2j$ brane on $X$, viewed in terms of the mirror
variety $Y$.  These periods will have branch cut discontinuities on the moduli space; only on the full Teichm\"{u}ller space are they continuous.  This is as the periods of the holomorphic $d$-form on a Calabi-Yau manifold
solve the generalized hypergeometric equation \cite{Teitelbaum,fuchsian,morrison_fuchs}:

\begin{equation}
\label{HG_eq} \left[ \delta~\prod_{i=1..q}{(\delta+\beta_i-1)}-
z~\prod_{j=1..p}{(\delta+\alpha_j)} \right]u=0~~. 
\end{equation}

\noindent where $\delta=z\partial_{z}$ and for a Calabi Yau of hypersurface type of dimension $d$, defined as the zero locus of a degree $d+2$ homogeneous polynomial in $\C\P^{d+1}$, we have $\alpha=\{\frac{i}{d+2}\}_{i=1}^{d+1}$ and $\beta =\{1\}_{j=1}^{d+1}$.  These model dependant constants are easily read off, in the case of complete intersections Calabi-Yau, by the divisor theory of the variety at hand, which determine the form of the regular solution to the above equation of Picard-Fuchs type.

The mirror variety $Y$ is defined as a quotient of:

\begin{equation} \label{eqdef}
\sum_{i=1}^{d+2}{x_{i}}^{(d+2)}-(d+2) \, \psi \, {x_{1}x_{2}\cdot\cdot\cdot x_{d+2}}=0
 \end{equation} 

\noindent by the identifications $x_i \simeq \omega^{k_i} x_i$, with $\omega=e^{
2\pi i/(d+2)}$ and the $k_i \in \Z$ satisfying $\sum_i k_i = 0$, where the $x_{i}$ 
are taken to be homogeneous coordinates on $\C \P^{d+1}$ and $\psi$ is a single
complex parameter which specifies the complex structure modulus of the 
mirror variety $Y$.  The $\psi$-plane is actually a $(d+2)$-fold covering of moduli
space, since $\psi$ and $\omega \psi$ yield isomorphic spaces
through the isomorphism $x_1 \to \omega x_1$; thus, $\psi$ is related to $z$ by $z=\psi^{(-d-2)}$.

A class of solutions to these PDE manifest themselves as Meijer
G-functions $U_j$, for $j\in\{0,d\}$ each with ${\rm log}^j z$ behavior around $z =
0$.  They possess the following integral representation:

\begin{equation}\label{intrep} \label{UJ} U_j(z)=\frac{1}{(2\pi
i)^j}\oint_{\gamma}\frac{ds
\Gamma(-s)^{j-1}\prod_{i=1}^{d}\Gamma(s+\alpha_i) ((-1)^{j+1}z)^s~
}{\Gamma(s+\beta_{j})^{d-j}}, \end{equation}

\noindent with $\alpha$ and $\beta$ as defined above.

This integral has poles at $\alpha_{i} -s = -n$ and
$\beta_{j}+s = -n$ for $n\in \Z^+$.  We may evaluate it by the
method of residues by choosing $\gamma$, a simple closed curve,
running from $-i\infty$ to $+i\infty$ in a path that separates the
two types of poles from one another.  Closing the contour $\gamma$
to the left or to the right will provide an asymptotic expansion
of $U_{j}(z)$ which is adapted to either the Gepner point
($\psi^{-(d+2)}=z=\infty$ in this parametrization) or the large
complex structure point ($z=0$).  Our choice of defining
polynomial for the varieties we study is such that the discriminant
locus (conifold point) lies at $z=1$.

One should note that the Meijer period expressions we write down are for the bare D-Brane masses, which have not been normalized through use of the K\"{a}hler potential.  A correctly normalized (and hence, non-holomorphic) mass equation is required to determine nonzero minima of $|Z(\Gamma)|$ for a given D-Brane charge $\Gamma$, since the minima of the norm of a holomorphic function is neccesarily zero.  However, for the question of vanishing cycles, we are able to proceed by considering only the bare mass.  It would be of interest to return to this class of higher dimensional examples with the neccesary integral structure at hand, to perform an analysis along the lines of \cite{spectrum}.

\subsection{Monodromies}

We quickly review the procedure by which one obtains the mondromy matrices for a given example, with the dimension $d$, and the sets $\alpha$ and $\beta$ as starting input.  The following is a brief summary of the exposition of \cite{GL1}; readers wishing greater detail should consult that reference.

The monodromy matrix $T[0]$ does not depend on the particular form of the coefficients $\alpha$, but only the form of $\beta$.  In our set of examples, $\beta = \{1,...,1\}$, and the matrix $T[0]$ only depends on the complex dimension $d$ of the variety in question (which is equal to the cardinality of $\beta$).

The canonical form of the monodromy matrix for the Gepner point, which we call $T_{can}[\infty]$, is derived from the general theory of partial differential equations of hypergeometric type, and is sensitive to the coefficients $\alpha$.  It is given by the following expression:

\begin{equation}
T_{can}[\infty]_{\mu\nu}=\delta_{\mu\nu}e^{-2\pi i \alpha_{\mu}}
\end{equation}

\noindent where $\mu$ and $\nu$ run from $0$ to $d$.  

The above can be brought, via a change of basis, into a form relevant to our chosen basis of periods $U^{\nu}$, by first conjugating with the following matrix $C$:

\begin{equation}
C_{\rho\sigma}=(\frac{\sin(\pi\alpha_{\sigma})}{\pi})^{d-\rho}(\delta_{\rho}^{even}+
e^{-i\pi\alpha_{\sigma}}\delta_{\rho}^{odd})
\end{equation}

\noindent and then with the following matrix $V$:

\begin{equation}
V_{\tau\omega}=\delta_{\tau\omega}(2\pi i)^{\tau}
\end{equation}

We then have $T[\infty]=VCT_{can}[\infty]C^{-1}V^{-1}$.  The conjugation by $V$  is needed to 
compensate for the normalization factor introduced in front of our integral representation of Meijer periods.  It also serves to greatly simplify the form of the monodromy matrices we list in the next section.

\subsection{Weakly Integral Periods}

At this point, the determination of a basis of weakly integral periods bears some review.  Since we focus our attention on examining D-branes wrapping cycles whose mass vanishes at singular points in the moduli space, we do not need to know exactly which linear combination of Meijer periods form a basis for an integral charge lattice.  We only need to find a basis which is proportonial as a whole, up to some nonzero complex number $\theta$, to such an integral basis.  Such a basis is called weakly integral in \cite{GL1}.  Should a linear combination of weakly integral periods have a zero somewhere in moduli space, then the (undetermined) integral class proportional to that period vector must also vanish at the same point.  

We generate a basis of weakly integral periods in $\theta H_3(Y,\Z)$ by finding first one weakly integral period, and then acting on it with monodromy transformations.  For hypersurface examples, the regular solution to the Hypergeometric equation is invariant under mondromy about the large complex structure point, and performing monodromies around the three regular singular points preserves the integral charge lattice of $H_3(Y,\Z)$ .  Thus, if the monodromy matrix $T[\infty]$ corresponding to the action on the periods after making a loop around the Gepner point is cyclic, i.e. there exists a positive integer $n>1$ such that $T[\infty]^n=1$, we may take the period vector with only one unit of $U_{0}$ charge and transport it around the Gepner point (n-1) times to generate a basis of $n$ weakly integral charge vectors.

\section{Collapsing Cycles}

\subsection{The Sextic Hypersurface}
\label{sec:sextic}
\setcounter{equation}{0}

We will first consider type IIA string theory compactified on the sextic hypersurface Calabi-Yau $X$, or equivalently type IIB compactified on its mirror $Y$, which is defined by a single homogeneous equation of Fermat type, as described above.  For the sextic, using the conventions detailed in \cite{GL1}, directly and straight forwardly generalized to a case of complex dimension four, and taking our periods to be defined in our normalization convention above, one can easily compute the monodromy matrices around these three regular singular points.  They are given by:

{\footnotesize \begin{equation}
\begin{array}{ccc}
T[0]=\left [\begin {array}{ccccc} 1&0&0&0&0\\-1&1&0&0&0\\1&-1&1&0&0\\0&0&-1&1&0\\0&0&1&-1&1\end {array}\right ]&~~~&
T[\infty]=\left [\begin {array}{ccccc} -5&4&-11&3&-6\\ -1&-1&0&0&0\\1&-1&1
&0&0\\ 0&0&-1 &1&0\\0&0&1&-1&1\end {array} \right ]\end{array} \end{equation}}

\noindent and $T[1]=T[\infty] \cdot T[0]^{-1}$ for $\im z < 0$,
$T[1]=T[0]^{-1} \cdot T[\infty]$ for $\im z > 0$.

We choose the period $U_0$, which is invariant under monodromy about the large complex structure point, as our generator.  Then we use the fact that $T[\infty]^6=1$ and transport $U_0$ by monodromy around the Gepner point, to obtain a weakly integral basis of periods $EU$ with the matrix $E$ given by:

\begin{equation}
E=\left [\begin {array}{ccccc} 1&0&0&0&0\\-5&4&-11&3&-6\\10&-5&35&-6&24\\--10&0&-45&0&-36\\5&5&29&6&24\\-1&-4&-8&-3&-6\end {array}\right ]
\end{equation}

A quick search indicates the following weakly integral period vanishes at the point $z=1$:

\begin{equation}
\Pi_{V}= 2U_{0}+4U_{1}+8U_{2}+3U_{3}+6U_{4}
\end{equation}

\vskip .25 in 

\hbox{\hskip 2 in \epsfysize 2 in \epsfbox{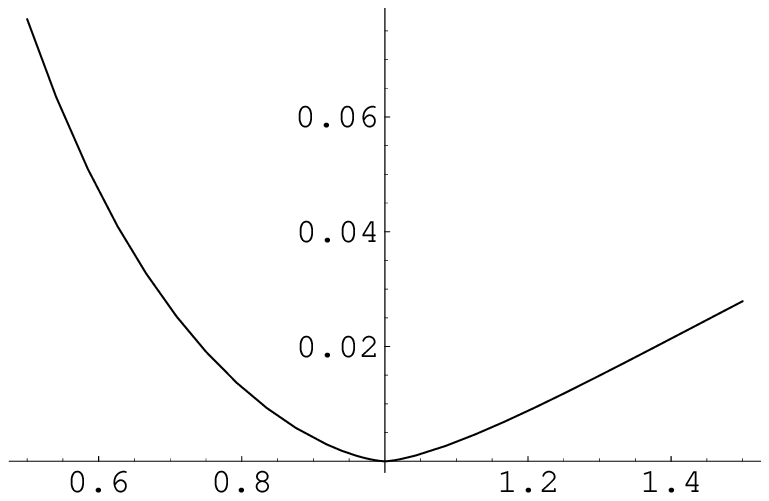}} 
\vskip .25 in 
\hbox{\centerline{\footnotesize Figure 1. Graph of $|Z(\Pi_{V})|$ around the point $z=1$  for the sextic hypersurface in $\C\P^5$.}}
\hbox{\centerline{\footnotesize The horizontal axis represents Re(z).}}
\vskip .25 in

\hbox{\hskip 1 in \epsfysize 2 in \epsfbox{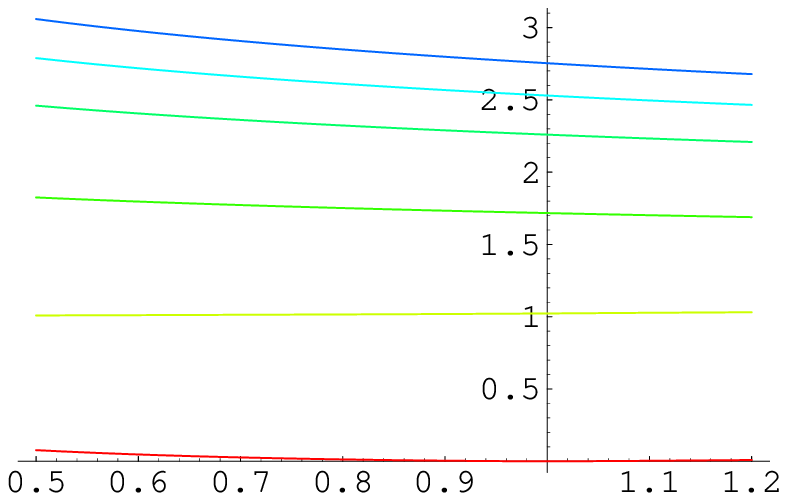}} 
\vskip .25 in 
\hbox{\centerline{\footnotesize Figure 2. Graph of $|U_{i}(z))|$ around the point $z=1$ for the sextic hypersurface in $\C\P^5$.}}
\hbox{\centerline{\footnotesize Included, in red, is $|Z(\Pi_{V})|$.  The horizontal axis represents Re(z).}}
\vskip .25 in

As indicated in the figures above, the pure n-dimensional periods $U_{i}(z)$ have a finite, nonzero, quantum volume around the point $z=1$, where the weakly integral period $\Pi_{V}$ vanishes.  Thus, in the case of this Calabi-Yau fourfold, we have a top dimensional cycle which can be seen as having zero quantum volume on the discriminant locus, as is the case with string theory compactified on a Calabi-Yau threefold such as the quintic.  However, the form of the vanishing cycle differs qualitatively from that of quintic hypersurface, having periods representing cycles of all dimensions.

\subsection{The Heptic Hypersurface}
\label{sec heptic}
\setcounter{equation}{0}

We next consider type IIA string theory compactified on the
heptic hypersurface Calabi-Yau $X$, or equivalently type IIB compactified on
its mirror $Y$, which is defined by a single homogeneous equation of Fermat type, as above.  For the heptic, using the conventions detailed in \cite{GL1}, directly and straight forwardly generalized to a case of complex dimension five,  and taking our periods to be defined as listed above one can easily compute the monodromy matrices around these three regular singular points.  They are given by:

{\footnotesize \begin{equation}
\begin{array}{ccc}
T[0]=\left [\begin {array}{cccccc} 1&0&0&0&0&0\\-1&1&0&0&0&0\\1&-1&1&0&0&0\\0&0&-1&1&0&0\\0&0&1&-1&1&0\\0&0&0&0&-1&1\end {array}\right ]&~~~&
T[\infty]=\left [\begin {array}{cccccc} 1&7&0&14&0&7\\ -1&-6&0&-14&0&7\\0&-1&1&0&0&0\\ 0&1&-1 &1&0&0\\0&0&0&-1&1&0\\0&0&0&1&-1&1\end {array} \right ]\end{array} \end{equation}}

\noindent and $T[1]=T[\infty] \cdot T[0]^{-1}$ for $\im z < 0$,
$T[1]=T[0]^{-1} \cdot T[\infty]$ for $\im z > 0$.

To generate a weakly integral basis of periods (i.e. a basis of periods which correspond to integral wrapped D-Branes, up to an overall complex normalization), we first chose the period $U_0$, which is invariant under monodromy about the large complex structure point.  We then transport $U_0$ by monodromy around the Gepner point to generate a weakly integral basis $EU$ with the matrix $E$ given by:

\begin{equation}
E=\left [\begin {array}{cccccc} 1&0&0&0&0&0\\1 &7&0&14&0&7\\-6&-21&-14&-63&-7&-35\\15&35&49&119&28&70\\-20&-35&-70&-119&-42&-70\\15&21&49&63&28&35\\-6&-7&-14&-14&-7&-7\end {array}\right ]
\end{equation}

We have a set of seven weakly integral periods because for the heptic hypersurface, $T[\infty]^7=1$.  Numerically evaluating these periods using an asymptotic expansion about the large complex structure point, one quickly finds the following linear combination of Meijer periods which vanishes at the conifold point $z=1$:

\begin{equation}
\Pi_{V}=U_{1}+2U_{3}+U_{5}
\end{equation}

\vskip .25 in 

\hbox{\hskip 2 in \epsfysize 2 in \epsfbox{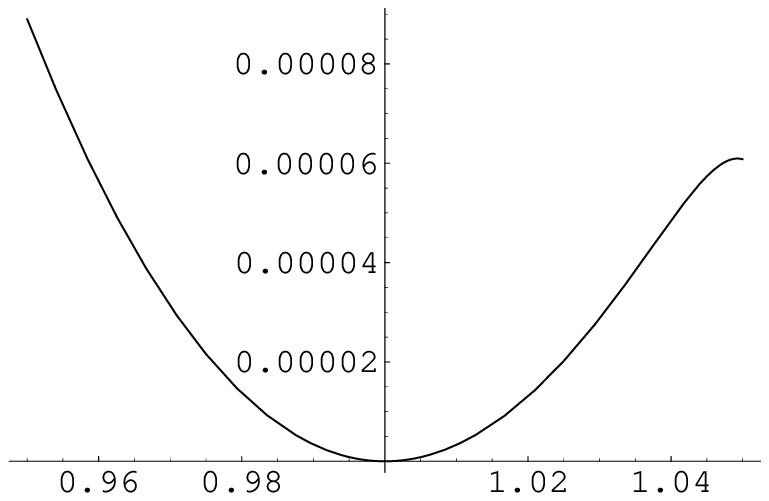}} 
\vskip .25 in 
\hbox{\centerline{\footnotesize Figure 3. Graph of $|Z(\Pi_{V})|$ around the point $z=1$ for the heptic example. }}
\hbox{\centerline {\footnotesize The horizontal axis represents Re(z).}}
\vskip .5 in 

\hbox{\hskip 1 in \epsfysize 2 in \epsfbox{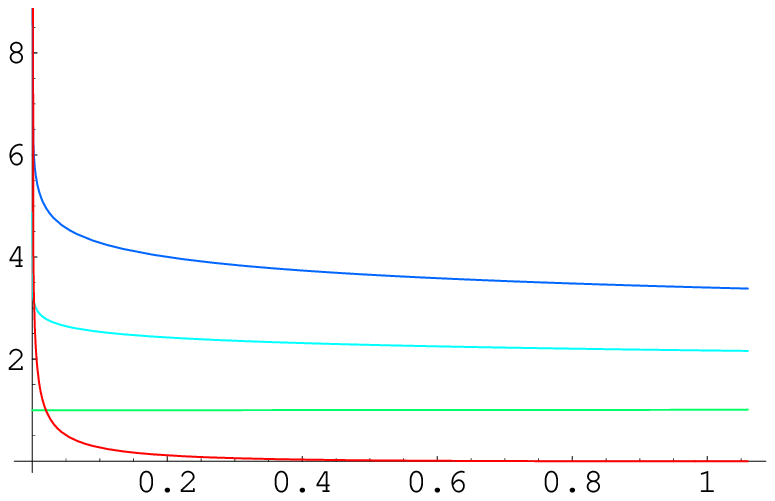}} 
\vskip .25 in 
\hbox{\centerline{\footnotesize Figure 4. Graph of $|U_{i}(z)|$ (for $i\in\{0,1,2\}$) around the point $z=1$ for the heptic example.}}
\hbox{\centerline {\footnotesize Included, in red, is $|Z(\Pi_{V})|$. The horizontal axis represents Re(z).}}
\vskip .5 in 

\hbox{\hskip 1 in \epsfysize 2 in \epsfbox{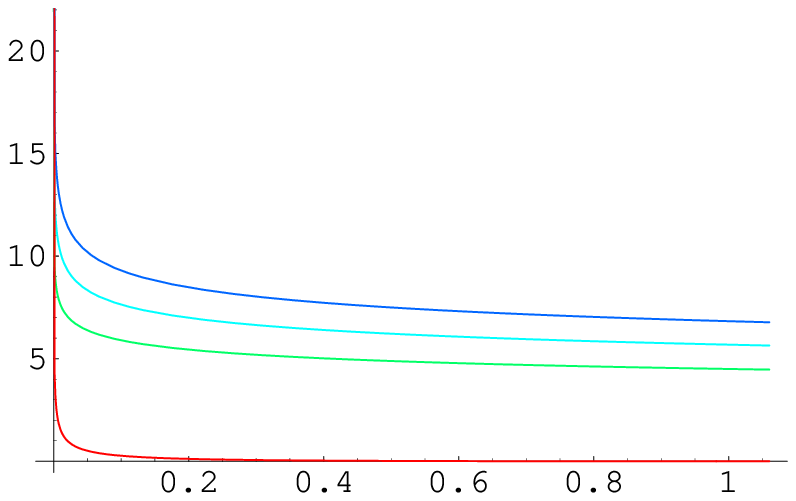}} 
\vskip .25 in 
\hbox{\centerline{\footnotesize Figure 5. Graph of $|U_{i}(z)|$ (for $i\in\{3,4,5\}$) around the point $z=1$ for the heptic example.}}
\hbox{\centerline {\footnotesize Included, in red, is $|Z(\Pi_{V})|$. The horizontal axis represents Re(z).}}
\vskip .25 in 

As indicated in the figures above, the pure n-dimensional cycles mirror to the periods $U_{i}(z)$ have a finite, nonzero, quantum volume around the point $z=1$, where the weakly integral period $\Pi_{V}$ vanishes.  Thus, Thus we have a top dimensional cycle with zero quantum volume on the discriminant locus.  In addition, its form is very similar to that of the quintic hypersurface, namely being a linear combination of periods which have leading logarithmic powers which are of odd dimension.

\subsection{A Weighted Example}
\label{sec:weighted}
\setcounter{equation}{0}

Finally, we consider type IIA string theory compactified on a hypersurface Calabi-Yau defined as the zero locus of a single homogeneous polynomial of degree 10 on the weighted projective space $\P(1,1,1,1,1,5)$.  The mirror variety can be described as an appropriate quotient of the zero locus of the following polynomial in the coordinates $x_{i}$ on the weighted space:

\begin{equation} \label{eqdef}
\sum_{i=1}^{5}{x_{i}}^{10}+x_{6}^{2}- \psi \, {x_{1}x_{2}\cdot\cdot\cdot x_{6}}=0
\end{equation} 

Hypersurfaces in weighted projective space obey similar hypergeometric equations to those in regular projective space, with a slight complication by the presence of the nontrivial weights.  However, one may still read off the period which is regular by examining the intersection theory of the variety, and thus determine the associated hypergeometric equation which the periods of the holomorphic fourform must satisfy, in exactly the same manner as the unweighted cases.  The equation one obtains is of the same form as listed above, with coefficients given by $\alpha=\{i/10\}_{i=1}^{9}$ and $\beta=\{1,1,1,1\}\cup\{\frac{j}{5}\}_{j=1}^{4}$.   In this example, the above coefficients to some extent cancel inside the integral expressions for the Meijer periods, leaving only $\alpha=\{1/10,3/10,1/2,7/10,9/10\}$ and $\beta=\{1,1,1,1\}$.  This makes the computation of residues straightforward.  Other, more general, examples may be studied; the only cost is a loss of computational simplicity, as generic $\beta$ values will introduce additional poles into the solutions and make the computation of the relevant monodromy matrices more cumbersome.  

For this weighted case, the monodromy matrices around the three regular singular points are:

{\footnotesize \begin{equation}\begin{array}{ccc}
T[0]=\left [\begin {array}{ccccc} 1&0&0&0&0\\-1&1&0&0&0\\1&-1&1&0&0\\0&0&-1&1&0\\0&0&1&-1&1
\end {array}\right ]&~~~&
T[\infty]=\left [\begin {array}{ccccc} -4&3&-7&1&-2\\ -1&-1&0&0&0\\1&-1&1
&0&0\\ 0&0&-1 &1&0\\0&0&1&-1&1\end {array} \right ]\end{array}\end{equation}}

\noindent and $T[1]=T[\infty] \cdot T[0]^{-1}$ for $\im z < 0$,
$T[1]=T[0]^{-1} \cdot T[\infty]$ for $\im z > 0$.

We choose the period $U_0$, which is invariant under monodromy about the large complex structure point, as our generator.  In this case, $T[\infty]^5=-1$, but we may still transport $U_0$ by monodromy around the Gepner point, to obtain a weakly integral basis of periods $EU$ with the matrix $E$ given by:

\begin{equation}
E=\left [\begin {array}{ccccc} 1&0&0&0&0\\-4&3&-7&1&-2\\6&-2&18&-1&6\\-4&-2&-17&-1&-6\\1&3&6&1&2\end {array}\right ]
\end{equation}

The following weakly integral period vanishes at the point $z=1$:

\begin{equation}
\Pi_{V}= 2U_{0}+3U_{1}+6U_{2}+1U_{3}+2U_{4}
\end{equation}

\vskip .25 in 

\hbox{\hskip 2 in \epsfysize 2 in \epsfbox{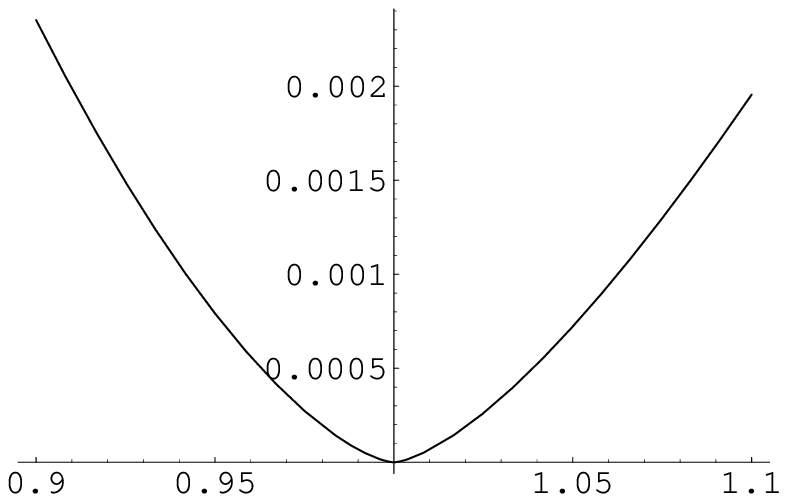}} 
\vskip .25 in 
\hbox{\centerline{\footnotesize Figure 6. Graph of $|Z(\Pi_{V})|$ around the point $z=1$  for the weighted example.}}
\hbox{\centerline{\footnotesize The horizontal axis represents Re(z).}}
\vskip .25 in

\hbox{\hskip 2 in \epsfysize 2 in \epsfbox{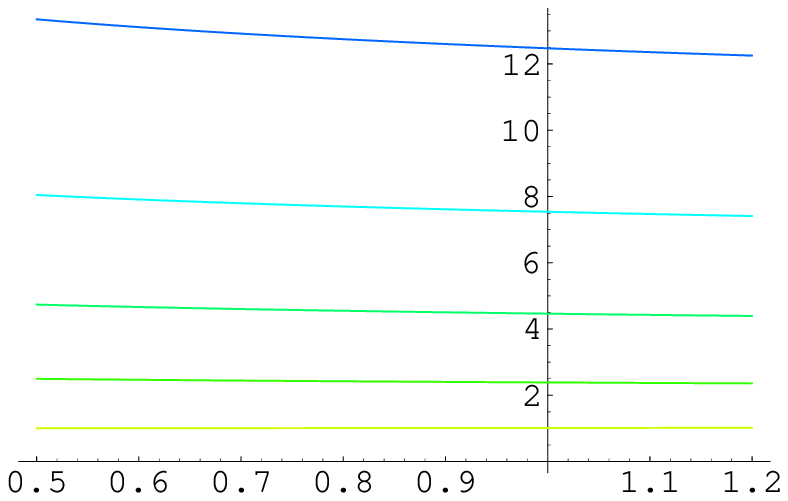}} 
\vskip .25 in 
\hbox{\centerline{\footnotesize Figure 7. Graph of $|U_{i}(z)|$ around the point $z=1$ for the weighted example.}}
\hbox{\centerline{\footnotesize The horizontal axis represents Re(z).}}
\vskip .25 in

As indicated in the figures above, the pure n-dimensional periods $U_{i}(z)$ have a finite, nonzero, quantum volume around the point $z=1$, where the weakly integral period $\Pi_{V}$ for this example vanishes.  The form of $\Pi_{V}$ in the weighted case is similar to that of the sextic, explicitly containing a mixture of cycles of all leading logarithmic behavior.

\section{Conclusions}

In this paper, we have extended the results of \cite{GL1} to include examples which occur in higher dimensions.  This is a natural direction to explore, and shows that even in higher dimensional Calabi-Yau manifolds, one may have a top-dimensional cycle with zero quantum volume, at a point in the moduli space where all pure cycles of smaller dimension remain at finite, nonzero, size. 

For fourfold examples, it might be of interest view to the manifolds in question as the internal part of a type IIB sting theory compactified down to two spacetime dimensions, and examine the relevant supergravity limit.  BPS black hole states corresponding to supersymmetric D-branes in the two extended spacetime directions should possess a simplified version of the attractor mechanism analogous to the one existing in four dimensions.  One might consider studying attractor flow lines in moduli space, and attempt to determine features of the spectrum of D-branes in the sextic and the weighted example, much in the same spirit as the analysis of \cite{spectrum}.  This is a topic to which we may hope to return to in future work.

\acknowledgments

The author would like to thank Brian Greene; this work was completed under his direction.  The author would also like to thank Gautham Chinta and Frederik Denef for useful discussions.


\begin{thebibliography}{99}

\bibitem{small_distances1}{P.~S.~Aspinwall, B.~R.~Greene, D.~R.~Morrison, 
 {\em  Measuring small distances in N=2 sigma models},
Nucl. Phys. {\bf B420} (1994) 184-242,\hepth{9311042}.}

\bibitem{quantum_volumes}{B.~R. Greene, Y.~Kanter,
 {\em   Small Volumes in Compactified String Theory},Nucl. Phys. 
{\bf B497} (1997) 127-145, \hepth{9612181}.}

\bibitem{morrison_quintic}{D.~R.~Morrison,
 {\em  Mirror symmetry and rational curves on quintic threefolds: a guide for 
mathematicians}, J. Amer. Math. Soc. {\bf 6} (1993) 223--247,alg-geom/9202004.}

\bibitem{D_geometry}{M.~R.~Douglas, {\em  Topics in D-geometry },
\hepth{9910170}.}

\bibitem{S}{A.~Strominger, {\em Massless Black Holes and Conifolds in String 
Theory}, Nucl.Phys. {\bf B451} (1995) 96-108, \hepth{9504090}.}

\bibitem{GL1}{B.~R.~Greene and C.~I.~Lazaroiu, \emph{Collapsing D-branes in
Calabi-Yau
moduli space, 1}, \hepth{0001025}},{~C.~I.~Lazaroiu, \emph{Collapsing D-branes in 
one-parameter models and small/large radius duality}, \hepth{0002004} .}

\bibitem{boundary_states}{ A.~Recknagel, V.~Schomerus,
 {\em  Moduli Spaces of D-branes in CFT-backgrounds},
\hepth{9903139}, 
{\em  Boundary Deformation Theory and Moduli Spaces of D-Branes},
Nucl.Phys. B545 (1999) 233-282, \hepth{9811237}, {\em  D-branes in Gepner models},
Nucl.Phys. B531 (1998) 185-225, \hepth{9712186}.}

\bibitem{Ishibashi}{N.~Ishibashi,  {\em  The boundary and crosscap states
 in conformal field theories}, Mod.~Phys.~Lett. {\bf A4} (1989) 251; 
 N.~Ishibashi, T.~Onogi,  {\em  Conformal field theories on surfaces
 with boundaries and crosscaps}, Mod.~Phys.~Lett. {\bf A4} (1989) 161.}

\bibitem{Douglas_quintic}{I.~Brunner, M.~R.~Douglas, A.~Lawrence, 
C.~Romelsberger, {\em  D-branes on the Quintic}, \hepth{9906200}.}   

\bibitem{Moore_arithmetics}{G.~Moore, 
{\em  Attractors and Arithmetic}, \hepth{9807056}, 
{\em Arithmetic and Attractors}, \hepth{9807087}.}

\bibitem{denef}{Frederik~Denef, \emph{Supergravity Flows and D-Brane Stability},
 \hepth{0005049}, \emph{On the correspondence between D-branes and stationary supergravity solutions of type II Calabi-Yau compactifications}, \hepth{0010222}.}

\bibitem{spectrum}{Frederik~Denef, Brian~R.~Greene, Mark~Raugas,
 \emph{Split Attractor Flows and the Spectrum of D-Branes on the Quintic}, \hepth{0101135}.}

\bibitem{higher}{Brian R. Greene, David R. Morrison, M. Ronen Plesser, \emph{Mirror Manifolds in Higher Dimensions}, \hepth{9402119}.}

\bibitem{Meijer_refs}{O.~I.~Marichev,
 {\em  Handbook of integral transforms of higher transcendental functions: 
theory and algorithmic tables},
Ellis Horwood series in mathematics and its applications, 
Halsted Press,New York 1983; 
A.~Erdelyi,W.~Magnus, F.~Oberhettinger,F.~G.~Tricomi,
 {\em  Higher transcendental functions};
Y.~Luke,
 {\em  The special functions and their approximations},Academic Press,1969.}

\bibitem{Norlund}{N.~E.~Norlund, {\em  Hypergeometric functions},
Acta Mathematica,{\bf 4} (1955),289--349.}

\bibitem{Teitelbaum}{A.~Libgober, J.~Teitelbaum, {\em  
Lines on Calabi-Yau complete intersections, mirror symmetry, and
Picard-Fuchs equations}, Internat. Math. Res. Notices 1993, {\bf no. 1}, 
29--39.}

\bibitem{fuchsian}{E.~A.~Coddington, N.~Levinson, {\em   Theory of ordinary 
differential equations}, New York, McGraw-Hill, 1955.}

\bibitem{morrison_fuchs}{D.~R.~Morrison,
 {\em  Picard-Fuchs equations and mirror maps for hypersurfaces},
in {\rm Essays on mirror manifolds}, 241--264, Internat. Press, 
Hong Kong, 1992, \hepth{9111025}.}

\bibitem{morrison_aspects}{ D.~R.~Morrison,  {\em  
Mathematical aspects of Mirror Symmetry}, Complex Algebraic Geometry 
(J. Kollar, ed.), IAS/Park City Math. Series, vol. 3, 1997, pp. 265-340, 
alg-geom/9609021.}

\end{thebibliography}
\end{document}